# 효율적인 양자화 기법을 통한 DNN 기반 화자 인식 모델 최적화

홍연아, 정우진, 강홍구

연세대학교 전기전자공학과

e-mail: [yeonahong, woojinchung]@dsp.yonsei.ac.kr, hgkang@yonsei.ac.kr

## Optimizing DNN-Based Speaker Verification: A Framework for Efficient Model Quantization

Yeona Hong, Woo-Jin Chung and Hong-Goo Kang

Department of Electrical and Electronic Engineering, Yonsei University

## Abstract

As Deep Neural Networks (DNNs) rapidly advance in various fields, including speech verification, they typically involve high computational costs and substantial memory consumption, which can be challenging to manage on mobile systems. Quantization of deep models offers a means to reduce both computational and memory expenses. Our research proposes an optimization framework for the quantization of the speaker verification model. By analyzing performance changes and model size reductions in each layer of a pre-trained speaker verification model, we have effectively minimized performance degradation while significantly reducing the model size. Our quantization algorithm is the first attempt to maintain the performance of the state-of-the-art pre-trained speaker verification model, ECAPA-TDNN, while significantly compressing its model size. Overall, our quantization approach resulted in reducing the model size by half, with an increase in EER limited to 0.07%.

## I. 서론

심층 신경망(DNN: Deep Neural Network) 모델 양자화(Quantization)[1] 기법은 학습된 원 모델의 기본 성능을 유지하면서도 전체 메모리 사용량 및 연산 속도를 줄이는 것을 목표로 한다. 인공 신경망 연구 초창기에는 하드웨어 성능이 제한적이고, 학습 데이터 등이 부족했기 때문에 다양한 종류의 인식 시스템의 모델 크기가 상대적으로 작을 수 밖에 없었고, 이에 따라 양자화 관련 연구는 큰 주목을 받지 못하였다. 하지만 하드웨어 가속기의 등장, 데이터 증가, 신경망 기술의 발전 등으로 더 큰 신경망 모델이 등장하고, 이에 따라 매우 복잡한 연산이 필요해 짐에 따라 기존의 실수형(floating-point) 연산기 보다 훨씬 메모리가 적게 필요하고, 연산이 빠른 16비트와 8비트 정수 연산기를 이용한 계산 방법에 대한 연구가 진행되었다[2-5]. 이미지 분류(image classification) 영역에서는 4개의 층을 가진 심층 신경망에 대해 2비트 양자화를 진행할 때, 18개의 층을 가진 심층 신경망을 4비트로 양자화 했을 때보다 분류 성능이 더 뛰어난 결과를 보인다는 양자화의 효율성을 언급한 연구 결과가 있다[6]. 또한, 1비트로 양자화된 대용량 언어 모델로도 충분한 성능을 얻을 수 있다는 것을 보여준 연구도 확인해 볼 수 있다[7].

본 논문은 컴퓨터 비전(computer vision)[8] 이나 자연어 처리(natural language processing)[9] 분야가 아닌 음성 분야에서도 양자화가 적용될 수 있음을 입증하고자 한다. 특히, 가장 최고의 성능(SOTA: state-of-the-art)을 보인다고 알려진 화자 인식 모델인 ECAPA-TDNN [10] (Emphasized Channel Attention, Propagation and Aggregation Time Delay Neural Network)에 다양한 양자화 기법을 적용하였고, 이를 통해 음성 신경망 모델에서

양자화를 적용함으로써 그 효율성을 검증하였다. ECAPA-TDNN은 가장 좋은 성능을 보임과 동시에 다른 화자 인식 모델에 비해 큰 모델 크기를 지니고 있으므로 양자화 적용 시 그 효율성을 강조하기에 적합하다 판단하였다. 또한, ECAPA-TDNN의 각 레이어 마다 독립적으로 양자화를 진행함으로써 모델 크기 감소 정도와 성능 하락 정도를 분석하였고, 이를 바탕으로 해당 신경망에 최적으로 적용 가능한 양자화 방식을 제안한다. 본 연구는 화자인식 성능 평가 척도인 EER (Equal Error Rate) 측면에서 단지 0.07% 저하된 성능을 유지하면서 신경망 모델의 크기를 53.44%로 줄이는 데 성공하였다.

## II. 본론

### 2.1 양자화

양자화는 매개변수와 계산을 나타내는 데 사용되는 숫자의 정밀도를 줄이는 과정을 의미한다. 심층 신경망에서의 양자화는 부동소수점 수를 사용하는 신경망을 낮은 비트를 필요로 하는 정수의 신경망으로 근사화 하는 과정으로서 메모리 저장 공간 및 연산에 필요한 복잡도를 모두 크게 감소시킨다. 이를 통해 휴대폰 및 임베디드 시스템과 같이 자원이 제한된 장치에서 더 빠른 추론과 낮은 전력의 소비를 가능하게 한다.

실수형 값인 $x$가 $x \in [\alpha, \beta]$ 사이에 포함되고 양자화된 $x_q$가 $x_q \in [\alpha_q, \beta_q]$ 사이에 포함될 때 양자화 수식은 다음과 같이 나타낼 수 있다.

$$x_q = round\left(\frac{x}{c} - d\right).$$

이 때, $c$는 스케일 값을 나타내며 양자화 이전의 부동소수점 값으로 변환할 때 사용하는 곱셈 계수이다. $d$는 영점으로 양자화된 값의 범위에서 원래의 값이 0에 대응하는 정수 값을 나타내게 된다. 이는 정수 값과 부동소수점 값 간의 중심을 맞추는 역할을 하게 된다. 양자화를 복원하는 식은 다음과 같이 나타낼 수 있다.

$$x = c(x_q + d).$$

양자화 과정에서 필요한 반올림 연산으로 인해 오차가 발생하게 되고, 해당 오차로 인해 양자화 이전 모델에 비해 성능이 하락할 수 있다.

양자화 기술은 훈련 후 양자화(PTQ: Post Training Quantization)와 양자화 인지 훈련 (QAT: Quantization Aware Training)[11]과 같이 크게 두 가지 방법으로 나눌 수 있다. 훈련 후 양자화는 이미 훈련이 완료된 모델이 가지고 있는 부동소수점 32비트의 연산을 정수형 16 혹은 8비트 연산으로 변환하여 추론 속도 향상 및 메모리 최적화를 달성하는 것이다. 이미 학습된 모델에 추론 과정에 사용될 데이터를 넣어 각 모델 변수의 크기 범위를 통계적으로 분석하여 가장 정보 손실 정도를 적게 할 수 있는 최댓값을 정하고, 이 밖에 범위는 버린다.

양자화 인식 학습은 학습 진행 시점에 수행 시 양자화 적용에 의한 영향을 미리 예상해 보는 방식으로 최적의 값을 구하는 것과 동시에 양자화를 하는 방식을 말한다. 훈련 중에 양자화의 영향을 미리 예상함으로써 정확도 감소의 폭을 최소화할 수 있는 장점이 있다.

두 방식은 훈련 속도나 정확도 등에서 차이가 있으므로 구현하고자 하는 시스템의 최종 목표에 맞춰 적합한 방법을 사용하는 것이 중요하다. 본 논문은 이미 학습되어 공개된 모델에 추가 학습 과정 없이 양자화 성능을 평가하는 것을 목표로 하므로 훈련 후 양자화 방식을 통하여 전체 실험 과정을 설계하고, 그 결과를 분석한다.

### 2.2 화자 인식

화자 인식은 음성 신호를 입력으로 하여 어떤 화자의 목소리인 지(speaker identification) 혹은 주장된 신원을 인증(speaker verification) 하는 것이다. 화자 인식은 사용의 편리함, 활용할 수 있는 다양한 애플리케이션[12,13], 심층 신경망 구조의 발전, 데이터 증가에 따라 최근 몇 년간 급격한 발전을 이루고 있다.

본 연구는 화자 인증 시스템을 목표로 하므로 다양한 평가 방법 중 EER (Equal Error Rate)을 성능 지표로 삼아 양자화 전후의 모델 성능 변화를 비교한다. EER은 오인식률(false acceptance rate)과 오거부율(false rejection rate)이 같아지는 임계점을 나타내는 것으로 낮을수록 좋다.

### 2.3 ECAPA-TDNN

ECAPA-TDNN은 화자 인식 시스템에서 뛰어난 성능을 발휘하는 최신 심층 신경망 구조로서 TDNN기반의 신경망을 개선하여 보다 효과적인 화자 인식을 가능하게 한다. 특히, Channel attention과 Global pooling을 통해 화자 인식의 정확성을 높인다.

ECAPA-TDNN은 Mel-spectrogram 또는 MFCC

(Mel Frequency Cepstral Coefficient)와 같은 음성의 음향적인 특징 벡터를 입력으로 받아 SE-Res2Block (squeeze-and-excitation residual network)을 통해 화자의 특징을 표현하는 latent embedding을 분석 구간 별로 추출한다. 이후 channel attention과 global pooling을 통해 고정 크기의 화자 embedding으로 만든다. 테스트 시에는 두 개의 음성을 입력으로 하여 학습된 모델에 통과시켜 각각 embedding을 추출한 후, 이를 비교하여 같은 화자인지 아닌지 판단하는 화자 인식을 수행하게 된다.

## Ⅲ. 구현

본 논문에서 화자인식 모델로 사용한 ECAPA-TDNN을 우선 Voxceleb2를 학습 데이터로 하여 RTX 3090 GPU에서 256 배치사이즈로 약 이틀 간 사전 학습한다. 또한, 성능 평가를 위해서는 ECAPA-TDNN 모델 학습 시 사용하지 않은 VoxCeleb1 [14] 데이터를 사용하여 양자화 전 후의 화자인식 성능 변화를 분석한다. VoxCeleb1 데이터는 40명의 화자가 녹음한 4715개의 음성 파일을 포함하고 있다.

양자화 과정의 일관성을 유지하기 위해 Pytorch-quantization toolkit을 이용하였고, 학습된 ECAPA-TDNN의 각 레이어 변수들에 대해 원하는 해상도의 양자화 비트로 모델링 한다. 양자화 과정을 통해 새롭게 모델링한 화자 인식 모델에 사전 학습된 모델 변수를 로드 후 교정 단계를 통해 기존의 변수를 원하는 비트 해상도로 연결한다. 교정 단계는 양자화에 있어서 가장 중요한 단계로 모델의 각 레이어에 대한 통계 데이터 (최댓값, 최솟값, 평균, 표준편차) 수집을 통해 각 레이어의 가중치와 활성화를 스케일링 하는 것이다. 스케일링이 완료된 후 이상이 없으면 교정된 가중치를 양자화된 모델과 함께 저장 후 사용한다. 최적화된 환경에서 양자화 모델을 수행하기 위해 본 연구에서는 엔비디아에서 제공하는 TensorRT [15] 환경에서 실험을 진행한다.

본 연구에서는 부동소수점 32비트로 구성된 신경망의 모든 레이어를 정수형 8비트로 양자화시킨 모델과 각 레이어 별로 양자화[16]를 적용했을 때의 성능을 비교 분석한다[17,18]. 이를 위해 각 레이어를 개별적으로 양자화 하는 모델링을 진행하였고, 변경된 모델을 바탕으로 실험을 통해 양자화 과정이 화자 인식 성능에 미치는 영향을 분석하고, 이를 통해 양자화의 효과를 극대화하면서도 성능 저하가 없는 최적의 모델을 선택하고자 한다. 실험 결과는 모델의 EER, 크기 등을 종합적으로 분석하여 판단한다.

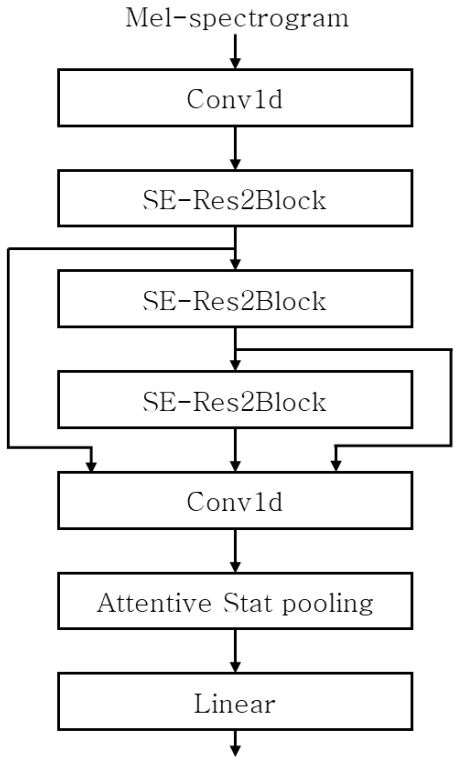

그림 1. ECAPA-TDNN 모델 구조

표 1. 레이어 별 양자화 결과

| Quantized layer | EER ↓ (%) | Model Size ↓ (MB) |
|---|---|---|
| No quantization | 1.665 | 63.571 |
| Conv1d | 1.660 | 60.413 |
| 1st SE-Res2Block | 1.680 | 53.501 |
| 2nd SE-Res2Block | 1.716 | 53.389 |
| 3rd SE-Res2Block | 1.766 | 53.381 |
| Conv1d | 1.688 | 48.365 |
| Attentive stat pooling | 1.681 | 56.594 |
| Linear | 1.665 | 60.100 |

표 2. 제안한 모델 성능 비교

| Quantized layer | EER ↓ (%) | Model Size ↓ (MB) |
|---|---|---|
| No quantization | 1.665 | 63.571 |
| Proposed | 1.739 | 33.974 |

## Ⅳ. 결과

표 1은 그림 1에 나타난 ECAPA-TDNN 모델의 레이어별 양자화 결과를 나타낸다. 표 1에서 SE-Res2Block들을 양자화 하였을 때의 성능 하락 추이를 통해 알 수 있듯이 SE-Res2Block이 ECAPA-TDNN의 최종 화자 인식 성능에 가장 큰 영향을 미치는 것을 확인할 수 있다. 더 나아가 SE-Res2Block들 중 첫 번째 레이어가 양자화에 대한 영향이 가장 적은 것을 확인할 수 있다. 이를 통해 화자 인식 모델의 시작 부분을 양자화 하는 것이 최종 인식이 이루어지는 뒷 쪽의 레이어를 양자화 하는 것에 비해 양자화 효과를 높일 수 있다는 것을 알 수 있다.

표 1의 결과를 바탕으로 본 연구에서는 두 번째 그리고 세 번째 SE-Res2Block을 제외한 나머지 레이어들을 양자화 하여 모델 크기 압축률 대비 성능 저하가 적은 모델을 제안한다. 그 결과 표 2에서 볼 수 있듯 양자화 이전 모델과 비교하여 단지 EER 0.07 % 증가만으로 최종 모델의 크기를 53.44%로 줄이는 결과를 얻을 수 있었다.

## Ⅳ. 결론 및 향후 연구 방향

본 논문에서는 컴퓨터 비전이나 자연어 처리 분야가 아닌 음성 분야에서도 인공 신경망 모델의 양자화를 효과적으로 적용할 수 있음을 입증하였다. 양자화 기술을 통해 ECAPA-TDNN 화자 인식 모델의 크기를 53.44%로 줄이는데 성공하였다. 특히, 모델의 각 레이어 양자화가 전체 성능에 미치는 영향을 분석하여 성능에 영향이 미미한 레이어 만을 양자화 함으로써 최종 성능 저하를 최소화하였다. 실험을 통해 양자화 기술을 화자 인식에 적용할 수 있고, 그리고 더 나아가 다른 심층 신경망 기반 음성 모델에 충분히 적용 가능함을 보였다. 이는 향후 온디바이스 음성 기반 어플리케이션의 효율성을 향상시킬 수 있는 기반이 될 것으로 보인다.

## 참고문헌